\documentclass[epj]{svjour}
\usepackage{graphics}
\usepackage{amssymb}
\usepackage{amsfonts}

\newcommand{\be}{\begin{eqnarray}}
 \newcommand{\ee}{\end{eqnarray}}
 \newcommand{\nee}{\nonumber\end{eqnarray}}
 \newcommand{\nn}{\nonumber\\ }
\begin{document}
\title{On Kaon production in  $e^+e^-$ and Semi-inclusive DIS
reactions}
\author{Ekaterina Christova\inst{1} \and Elliot Leader\inst{2}}% etc

\institute{Institute for Nuclear Research and Nuclear Energy,
        Sofia 1784, Bulgaria; echristo@inrne.bas.bg \and Imperial College, London, UK; e.leader@imperial.ac.uk}
\date{Received: date / Revised version: date}
\abstract{ We consider semi-inclusive unpolarized DIS for the
production of charged kaons and the different possibilities
 to test the conventionally used assumptions $s-\bar
s=0$ and $D_d^{K^+-K^-}=0$. The considered tests have the
advantage that they do not require
  any knowledge of the fragmentation functions.
We also show that measurements of both
 charged and neutral kaons would allow the determination of the kaon FFs
$D_q^{K^++K^-}$ solely from SIDIS measurements, and discuss the
comparison of $(D_u-D_d)^{K^+-K^-}$ obtained independently in
SIDIS and $e^+e^-$ reactions. All analysis are performed in LO and NLO in QCD.
The feasibility of the tests to HERMES SIDIS data is considered.}

\maketitle
%
%%%%%%%%%%%%%%%%%%%%%%%%%%%%%%%%%%%%%%%%%%%%%%%%%%%%%%%%%%%%%%%%%%%%%%%%%%%%%55
\section{ Introduction}
\label{intro}
%%%%%%%%%%%%%%%%%%%%%%%%%%%%%%%%%%%%%%%%%%%%%%%%%%%%%%%%%%%%%%%%%%%%%%%%%%%%%%%%%555
It is well known that neutral current inclusive deep inelastic
scattering (DIS) yields information only about quark plus
antiquark parton densities. When neutrino experiments are possible
one can obtain separate knowledge about quark and antiquark
densities, but for the case of polarized DIS this is impossible
experimentally. For this case semi-inclusive DIS (SIDIS), where
some final hadron is detected, plays an essential role, but
requires a knowledge of the fragmentation function (FF) for a
given parton to fragment into the relevant hadron. As pointed out
in \cite{Kretzer_we} and more recently in  \cite{deFlorian} a precise
knowledge of the FFs is vital. In this paper we examine what we can learn
about the kaon FFs  from experimental data.

When the spin state of the detected hadron is not monitored, it is
possible to learn about the FFs from both $e^+e^-\rightarrow hX $
and \textit{unpolarized} SIDIS $ l+N \rightarrow l hX $. In the
case of pion production $SU(2)$ plays a very helpful role in
reducing the number of independent FFs needed. For
production of charged kaons, which is important for studying the strange quark
densities, $SU(2)$ is less helpful, and even a combined analysis
of $e^+e^-$ and SIDIS data on both protons and neutrons does not
allow an unambiguous determination of the kaon FFs \cite{Dubna05}.

It is thus conventional to make certain reasonable sounding
assumptions about the strange quark densities and the kaon FFs. In
the first part of this paper we discuss to what extend these assumptions can be
justified and tested experimentally.  We shall discuss tests based on both a
leading order (LO) and a next-to-leading order (NLO)
approach. For although it has been often assumed that an NLO treatment
is essential, in our paper we have kept the LO treatment for two reasons --
one always starts with LO and then follows the natural hierarchy LO $\to$ NLO and also because
a recent study in \cite{deFlorian} showed that a
very acceptable description of the combined polarized DIS and SIDIS data  can be achieved in a LO
 approximation as well and thus LO cannot be ruled out yet.

As mentioned above, SU(2)symmetry is of little help if only charged kaons are measured.
However, it is well known that charged and neutral kaons are combined in SU(2) doublets.
This relates the FFs of $K_s^0$ to those of $K^\pm$, which implies that no new FFs appear in $K_s^0$-production.
In the second part of our paper we examine to what extend detecting neutral  as well as charged
kaons can help to determine the kaon fragmentation functions. We carry out the analysis in LO and NLO.

In Section 2 we  recall the
general formulae for inclusive $e^+e^-$ and SIDIS. In Section 3 we
consider semi-inclusive $K^\pm$ production and  possible tests
whether, for the quark densities, $s(x) =\bar s(x) $, and whether,
for the fragmentation functions,  $ D_d^{K^+}(z) =D_d^{K^-}(z)$.
In Section 4 we discuss production of $K^\pm ,\, K^0_s$; in Sections 5 and 6 we consider
 the combinations $K^+ +K^-- 2K^0_s$ and  $K^+ +K^-+ 2K^0_s$ respectively, both in LO and NLO.
 Possible tests for the reliability of the leading order
treatment of the processes are discussed.

%%%%%%%%%%%%%%%%%%%%%%%%%%%%%%%%%%%%%%%%%%%%%%%%%%%%%%%%%%%%%%%%%%%%%%%%%%%%%55
\section{ General formula for  $e^+e^-$ and unpolarized SIDIS}
\label{sec:1}
%%%%%%%%%%%%%%%%%%%%%%%%%%%%%%%%%%%%%%%%%%%%%%%%%%%%%%%%%%%%%%%%%%%%%%%%%%%%%%%555

For convenience we shall recall some general formulae for the
cross sections and asymmetries in $e^+e^-\to hX$ and $e+N\to
e+h+X$.

 %%%%%%%%%%%%%%%%%%%%%%%%%%%%%%%%%%%%%%%%%%%%%%%%%%%%%%%%%%%%%%%%%%%%%%%%%%%%%55
\subsection{  $e^+e^-\to hX$ }
\label{sec:2}
%%%%%%%%%%%%%%%%%%%%%%%%%%%%%%%%%%%%%%%%%%%%%%%%%%%%%%%%%%%%%%%%%%%%%%%%%%%%%%%%%555

There are two distinct measurements of interest : the total cross
section $d\,\sigma_T^h(z)$ and the forward backward (FB) asymmetry
$A^h_{FB}$.  If $d^2\sigma^h/(dz\,d\,\cos\theta )$ is the
differential cross section for $e^+e^- \to hX$, these quantities
are defined as: \be
d\,\sigma_T^h(z)&=&\int_{-1}^{+1} \,\left(\frac{d^2\sigma^h}{dz\,d\,\cos\theta}\right)\, d\cos\theta\\
A^h_{FB}(z)&=&\left[\int_{-1}^{0}-\int_{0}^{+1}\right]\,\left(\frac{d^2\sigma^h}{dz\,d\,\cos\theta}\right)\,
d\cos\theta , \ee
 where $\theta $ is the CM scattering angle and
$z$ is,   neglecting masses,  the fraction of the  momentum of the
fragmenting parton transferred to the hadron $h$: $z=
2(P^h.q)/q^2=E^h/E $, where $E^h$ and $E$ are the CM energies of
the final hadron $h$ and the initial lepton.

  From  CP invariance it follows that
  \be
d\,\sigma_T^h(z)=d\,\sigma_T^{\bar h}(z),\qquad
A^h_{FB}(z)=\,-\,A^{\bar h}_{FB}(z),\label{CP} \ee where $\bar h$
is the C-conjugate of the hadron $h$. Eq. (\ref{CP}) implies that
 the total cross section $d\,\sigma_T^h$ actually provides information only about
  $D_q^{h+\bar h}\equiv D_q^{h} + D_q^{\bar h}$, while measurement
  of $ A^h_{FB}$  determines the non-singlet   (NS)
  combinations $D_q^{h-\bar h}\equiv D_q^{h} - D_q^{\bar h} $,
  and this is true in all orders of QCD.

In LO the formula are especially simple:
\be
d\sigma^h_T(z)&=&
3\,\sigma_0\,\sum_q \hat e_q^2\,D_q^{h+\bar h},\quad \sigma_0=\frac{4\pi\alpha_{em}^2}{3\,s}\\
A_{FB}^h(z)&=&3\,\sigma_0\,\sum_q \frac{3}{2}\,\,\hat
a_q\,D_q^{h-\bar h}.
\ee
Assuming both photon  and $Z^0$-boson
exchange we have:
\be
\hat{e_q}^2(s) &=&e_q^2 - 2e_q
\,v_e\,v_q\,\Re e \,h_Z +\nn
&&+ (v_e^2 + a_e^2) \, \left[(v_q)^2
+(a_q)^2\right]\, \vert h_Z\vert ^2\nn
\hat a_q &=&
2\,a_e\,a_q\,\left(-e_q\, \Re e\,h_Z + 2\,v_e\,v_q\,\vert
h_Z\vert^2\right),
\ee
 where $ h_Z =
[s/(s-m_Z^2+im_Z\Gamma_Z)]/\sin ^2 2\theta_W$.
 In (6) $e_q$ is the
 charge of the quark $q$ in units of the proton charge, and, as
  usual,
 \be
 v_e&=&-1/2 +2 \sin^2\theta_W,\quad a_e=-1/2, \nn
v_q&=&I_3^q-2e_q\sin^2\theta_W,\quad a_q=I_3^q,\nn
&&\qquad \quad I_3^u = 1/2,
\quad I_3^d = -1/2.
\ee

%%%%%%%%%%%%%%%%%%%%%%%%%%%%%%%%%%%%%%%%%%%%%%%%%%%%%%%%%%%%%%%%%%%%%%%%%%%%%%%%%%%
\subsection{ Unpolarized  SIDIS $e+N\to e+h+X$ }
\label{sec:3}
%%%%%%%%%%%%%%%%%%%%%%%%%%%%%%%%%%%%%%%%%%%%%%%%%%%%%%%%%%%%%%%%%%%%%%%%%%%%%%%%%%%%%%

 In semi-inclusive deep inelastic scattering, we consider the non-singlet
 difference
of cross-sections
 $\sigma_N^{h-\bar h}$,  where the measurable quantity is the ratio:
 \be
R_N^{h-\bar h}=\frac{ \sigma_N^{h-\bar h}}{\sigma_N^{DIS}},\qquad
\sigma_N^{h-\bar h}= \sigma_N^{h}-\sigma_N^{\bar h}. \ee
 For simplicity, we use  $\tilde\sigma_N^h$ and
$\tilde\sigma_N^{DIS}$ in which common kinematic factors have been
removed:
 \be
&&\tilde\sigma_N^h \equiv \frac{x(P+l)^2}{4\pi \alpha^2} \left( \frac{2y^2}{1+(1-y)^2}\right)\frac{d^3\sigma^h_N}{dxdydz}\\
&&\tilde\sigma_N^{DIS} \equiv \frac{x(P+l)^2}{4\pi \alpha^2}
\left(
\frac{2y^2}{1+(1-y)^2}\right)\frac{d^2\sigma^{DIS}_N}{dxdy}.
\ee
 Here $P$ and $l$ are the nucleon and lepton four momenta, and
 $x,y,z$ are the usual deep inelastic kinematic variables: $ x= Q^2/2P.q =Q^2/2M\nu $,  $ y= P.q/P.l =\nu /E$,
   $ z= P.P_h/P.q= E^h/\nu$, where
  E and $E^h$ are the lab. energies of the incoming lepton and final hadron.
 Note that, both in $e^+e^-$ and in SIDIS, neglecting masses,  $z$   always
 measures the fraction of the  parton momentum transferred to the produced hadron.

  Since the kinematic factors for $\sigma_N^h$ and $\sigma_N^{DIS}$ are the same, we can write:
\be
\tilde\sigma_N^{h-\bar h}=R_N^{h-\bar h}\tilde\sigma_N^{DIS},
\ee
where for $\tilde\sigma_N^{DIS}$ any of the parametrizations for the structure functions
$F_2$ and $R$  or, equivalently, any set of the unpolarized parton densities (PD)
can be used.

As shown in  \cite{Dubna05}, the general expression for the  cross
section differences, in NLO, is given by:
 \be
\tilde \sigma_p^{h-\bar h}(x,z)&=&\frac{1}{9}\left[4
u_V\otimes D_u^{h-\bar h} +  d_V\otimes D_d^{h-\bar h} \right.\nn
&& \left.+ (s -\bar
s)\otimes D_s^{h-\bar h}\right]\otimes  \hat\sigma_{qq} (\gamma q
\to q X),\nn
 \tilde \sigma_n^{h-\bar h}(x,z)
&=&\frac{1}{9}\left[4 d_V\otimes D_u^{h-\bar h} +  u_V\otimes
D_d^{h-\bar h} \right.\nn
&&\left. + (s -\bar s)\otimes D_s^{h-\bar h}\right]\otimes
\hat\sigma_{qq} (\gamma q \to q X). \label{diff}
 \ee
Here $\hat\sigma_{qq}$ is  the perturbatively  calculable, hard
partonic cross section $q\gamma^*\to q+X$:
\be
 \hat\sigma_{qq} &=&  \hat\sigma_{qq}^{(0)} +
 \frac{\alpha_s}{2\pi} \hat\sigma_{qq}^{(1)}\,,
 \ee
normalized so that $ \hat\sigma_{qq}^{(0)} = 1 $.

It is seen that  $ \tilde \sigma_N^{h-\bar h}$ involves only  NS
parton densities and fragmentation functions, implying that its
$Q^2$ evolution is relatively simple.
 Eq.(\ref{diff})  is sensitive to the valence quark
densities,  but also to the completely unknown combination
$(s-\bar s)$. The term $(s-\bar s)D_s^{h-\bar h}$ plays no role in
Pion production, since, by SU(2) invariance,
$D_s^{\pi^+-\pi^-}=0$. However it is important for kaon
production, for which $D_s^{K^+-K^-}$ is a favoured transition,
and thus expected to be big.

Up to now all analyses of experimental data have assumed $s=\bar
s$. In the next Sections we shall consider  the production of
kaons, $h=K^\pm$ and  $h=K^\pm , K^0_s$ and show how this
assumption,  and the assumption $D_d^{K^+-K^-}=0$, can be tested
without requiring knowledge of the FFs.

%%%%%%%%%%%%%%%%%%%%%%%%%%%%%%%%%%%%%%%%%%%%%%%%%%%%%%%%%%%%%%%%%%%%%%%%%%%%%%%%%%%
\section{Production of charged kaons }
\label{sec:4}
%%%%%%%%%%%%%%%%%%%%%%%%%%%%%%%%%%%%%%%%%%%%%%%%%%%%%%%%%%%%%%%%%%%%%%%%%%%%%%%%%%%%%%

%\vspace{.5cm}

As seen from (\ref{diff}), in $R_N^{K^+-K^-}$ both $s-\bar s$ and
$D_d^{K^+-K^-}$ appear. They are expected to be small, and the
usual assumption is that they are equal to zero. Here we examine
to what extent one can test these assumptions experimentally in
SIDIS.

It was shown in   \cite{Dubna05}, that even if we combine   data
on the forward-backward asymmetry $A_{FB}^{K^+-K^-}$ in
$e^+e^-$-annihilation with measurements of $K^+$ and $K^-$
production in SIDIS, we cannot determine the fragmentation
functions without   assumptions. The reason is that
 we have 3 measurements for the 4 unknown quantities
$D_{u,d,s}^{K^+-K^-}$ and ($s-\bar s$). Thus, one needs an
assumption:   either $s-\bar s=0$ or $D_d^{K^+-K^-}=0$. In fact, up
to now, all analyses of experimental data have been performed
assuming both $s-\bar s=0$ and $D_d^{K^+-K^-}=0$.

 Note, that from the quark content of $K^\pm$, the assumption
$D_d^{K^+-K^-}=0$ seems very reasonable if the $K^\pm$ are
directly produced. However, if they are partly produced via
resonance decay this argument is less persuasive. Of course
$e^+e^-\to K^\pm X$ sheds no light on this issue.

%%%%%%%%%%%%%%%%%%%%%%%%%%%%%%%%%%%%%%%%%%%%%%%%%%%%%%%%%%%%%%%%%%%%%%%%%%%%%%%%%%%%%%%%%%%%%%%%%%
\subsection{ LO approximation, $K^\pm$}
\label{sec:5}
%%%%%%%%%%%%%%%%%%%%%%%%%%%%%%%%%%%%%%%%%%%%%%%%%%%%%%%%%%%%%%%%%%%%%%%%%%%%%%%%%%%%%%%%%%%%%%%%%%

 In {\bf LO} we have:
\be
\tilde\sigma_p^{K^+-K^-} &=& \frac{1}{9}
[4\,u_V\,D_u^{K^+-K^-} +
d_V\,D_d^{K^+-K^-} \nn
&&+(s-\bar s)\,D_s^{K^+-K^-}],\\
\tilde\sigma_n^{K^+-K^-} &=& \frac{1}{9} [4\,d_V\,D_u^{K^+-K^-} +
u_V\,D_d^{K^+-K^-} \nn
&& +(s-\bar s)\,D_s^{K^+-K^-}].
\ee

From a theoretical point of view it is more useful to consider the
following combinations of cross-sections, which, despite involving
 differences of cross-sections, are likely to be large:
 \be
 (\tilde\sigma_p - \tilde\sigma_n)^{K^+-K^-}&=&\frac{1}{9} [ (u_V-d_V)\,(4D_u-D_d)^{K^+-K^-}],\nn
 (\tilde\sigma_p +\tilde\sigma_n)^{K^+-K^-}&=&\frac{1}{9} [
(u_V+d_V)\,(4D_u+D_d)^{K^+-K^-}\nn
&&\quad +2(s-\bar s)D_s^{K^+-K^-} ].
\ee
We
define:
\be
R_+(x,z)&\equiv&\frac{(\tilde\sigma_p +
\tilde\sigma_n)^{K^+-K^-}}{u_V+d_V},\nn
R_-(x,z)&\equiv&\frac{(\tilde\sigma_p -
\tilde\sigma_n)^{K^+-K^-}}{u_V-d_V}.
\ee

From a study of the $x$ and $z$ dependence of these we can deduce
the following:

1) if  $R_-(x,z)$ is a function of $z$ only, then  this suggests
that a LO approximation is reasonable.

2) if $R_+(x,z)$ is  \textit{also} a function of $z$ only, then,
since $D_s^{K^+ - K^-} $ is a favoured transition, we can conclude
that $(s-\bar s)=0$.

 3) if $R_+(x,z)$ and $R_-(x,z)$ are \textit{both} functions of $z$ only, and if in addition,
 $R_+(x,z) = R_-(x,z)$, then both \\$s-\bar s=0$
{\it and} $D_d^{K^+-K^-}= 0$.

4) if $R_+(x,z)$ and $R_-(x,z)$ are \textit{both} functions of $z$
only, but they are \textit{not} equal, $R_+(x,z) \neq R_-(x,z)$,
we conclude that $s-\bar s=0$, {\it but} $D_d^{K^+-K^-}\neq 0$.

5) if $R_-(x,z)$ is not a function of $z$ only, then NLO
corrections are needed, which we  consider below.

The above tests for $s-\bar s =0$ and $D_d^{K^+-K^-}= 0$ can be spoilt either by
$s-\bar s \neq 0$ and/or $D_d^{K^+-K^-}\neq 0$, or by NLO corrections,
which are both complementary in size. That's why it is important to formulate
tests sensitive to $s-\bar s =0$ and/or $D_d^{K^+-K^-}= 0$ solely, i.e. to consider NLO.

%%%%%%%%%%%%%%%%%%%%%%%%%%%%%%%%%%%%%%%%%%%%%%%%%%%%%%%%%%%%%%%%%%%%%%%%%%%%%%%%%%%%%%%%%%%%%%%%%%
\subsection{NLO approximation, $K^\pm$}
\label{sec:6}
%%%%%%%%%%%%%%%%%%%%%%%%%%%%%%%%%%%%%%%%%%%%%%%%%%%%%%%%%%%%%%%%%%%%%%%%%%%%%%%%%%%%%%%%%%%%%%%%%%

In an NLO treatment  it is still possible to reach
some conclusions, though less detailed than in the LO case. We now
have:
 \be
&&\qquad\qquad (\tilde\sigma_p - \tilde\sigma_n)^{K^+-K^-}=\nn
&&= \frac{1}{9}
(u_V-d_V)
\otimes(1+\alpha_s\,{\cal C}_{qq})\otimes  (4D_u-D_d)^{K^+-K^-}\label{NLOp}\\
&&\qquad\qquad (\tilde\sigma_p + \tilde\sigma_n)^{K^+-K^-}=\nn
&&= \frac{1}{9}
\left[(u_V+d_V)\,\otimes\, (4D_u+D_d)^{K^+-K^-}
+\right.\nn
&&\qquad\qquad \left.2(s-\bar
s)\,\otimes\,D_s^{K^+-K^-}\right]\,\otimes\,(1+\alpha_s\,{\cal C}_{qq}).\label{NLOn}
\ee
Here ${\cal C}_{ij}$ are
\be
{\cal C}_{ij}(y)&=&C_{ij}^M+[1+4\gamma (y)]C_{ij}^L\nn
\gamma (y) &=&\frac{1-y}{1+(1-y)^2}
\ee
 $C_{ij}^{M,L}$ being the corresponding Wilson coefficients \cite{Vogelsang}.
Suppose we try to fit  both (\ref{NLOp}) and (\ref{NLOn}) with one
and the same fragmentation function $D(z)$:
\be
&&\qquad\qquad (\tilde\sigma_p - \tilde\sigma_n)^{K^+-K^-}\approx\nn
&&\,\,\approx \frac{4}{9} (u_V-d_V)\,\otimes\,(1+\alpha_s\,{\cal C}_{qq})\otimes\,D(z),\label{NLO1}\\
&&\qquad\qquad (\tilde\sigma_p + \tilde\sigma_n)^{K^+-K^-}\approx\nn
&&\,\,\approx \frac{4}{9}
(u_V+d_V)\,\,\otimes\,(1+\alpha_s\,{\cal C}_{qq})\otimes\,D(z).\label{NLO2}
\ee

If this gives an acceptable fit for the $x$ and $z$-dependence of both $p-n$ and $p+n$ data, we can
conclude that both
$s-\bar s \approx 0$ {\it and} $D_d^{K^+-K^-}\approx 0$ and that $D(z) = D_u^{K^+-K^-}$.

Note that for all above tests, both in LO and NLO approximation,
we don't require a knowledge of $D_{u,d}^{K^+-K^-}$. This is
especially important since the  $e^+e^-$ total cross section data
determine only the $D_{q}^{K^++K^-}$, and these are relatively
well known, while $D_{u,d}^{K^+-K^-}$ can be determined solely
from $A_{FB}$ in $e^+e^-$  or from SIDIS.

The results of the above tests would indicate what assumptions are
reliable in trying to extract the   fragmentation functions
$D_{u,d,s}^{K^\pm}$ from the same data.

%%%%%%%%%%%%%%%%%%%%%%%%%%%%%%%%%%%%%%%%%%%%%%%%%%%%%%%%%%%%%%%%%%%%%%%%%%%%%%%%%%%%%%%%%%%%%%%%%%
\section{ Production of charged and neutral kaons}
\label{sec:7}
%%%%%%%%%%%%%%%%%%%%%%%%%%%%%%%%%%%%%%%%%%%%%%%%%%%%%%%%%%%%%%%%%%%%%%%%%%%%%%%%%%%%%%%%%%%%%%%%%%

 The description of SIDIS and $e^+e^-$  reactions, in which one monitors neutral
  $K_s^0=(K^0 +\bar K^0)/\sqrt 2$ as well as
charged $K^\pm$ does not require any further FFs.
 This is due to SU(2)  invariance which  relates the neutral to the charged kaon FFs:
\be
 D_u^{K^+ + K^-}&=&D_d^{K^0+ \bar K^0},\qquad
D_d^{K^+ + K^-}=D_u^{K^0+ \bar K^0},\nn
 D_s^{K^+ + K^-}&=&D_s^{K^0+ \bar K^0},\qquad D_g^{K^+ + K^-}=D_g^{K^0+ \bar K^0}.\label{SU2}
\ee
In principle this helps to determine the kaon FFs
$D_{u,d,s}^{K^++K^-}$ solely from SIDIS measurements, without the
problem of combining $e^+e^-$ data and SIDIS data at widely
different value of $Q^2$.

 Two possible measurements can be performed:
with   ($K^+ +K^--  2\,K^0_s$) and with ($K^+ +K^-+  2\,K^0_s$).

%%%%%%%%%%%%%%%%%%%%%%%%%%%%%%%%%%%%%%%%%%%%%%%%%%%%%%%%%%%%%%%%%%%%%%%%%%%%%%%%%%%%%%%%%%%%%%%%%%
\section{The combination $K^+ +K^-- 2K^0_s$}
\label{sec:8}
%%%%%%%%%%%%%%%%%%%%%%%%%%%%%%%%%%%%%%%%%%%%%%%%%%%%%%%%%%%%%%%%%%%%%%%%%%%%%%%%%%%%%%%%%%%%%%%%%%

In NLO we have

for $e^+e^-$:
\be
&&\qquad\qquad d\sigma_T^{K^+ +K^-- 2K^0_s}(z)=\nn
&&=3\,\sigma_0\,(\hat e^2_u - \hat e^2_d )_{m_Z^2} \,\left[1+\frac{\alpha_s}{2\pi}C_F\,(c_T^q+c_L^q)\otimes \right]\nn
&&\qquad\qquad\qquad \times (D_u - D_d)^{K^+ +K^-},\label{a}\\
&&\quad d\sigma^{K^+ +K^-- 2K^0_s}_T \equiv
d\sigma^{K^+}_T\,+d\sigma^{K^-}_T\,-2\,d\sigma^{K^0_s}_T,
 \nee
where  $c^q_{T,L}$ are the Wilson coefficients for the contribution of
the transverse (T) and longitudinal (L) virtual boson \cite{Kretzer}.

and for SIDIS:
\be
&&\qquad\qquad \tilde\sigma_p^{K^+ +K^-- 2K^0_s}(x,,y,z)=\nn
&&=\left\{\frac{1}{9} \,[4(u+\bar u) -(d+\bar d)]\left(1+\frac{\alpha_s}{2\pi}\otimes {\cal C}_{qq}\otimes\right) +\right.\nn
&&\qquad\left.+\frac{1}{3}\frac{\alpha_s}{2\pi} g\otimes {\cal C}_{gq}\otimes\right\}(D_u - D_d)^{K^+ +K^-},\label{b}\\
&&\qquad\qquad \tilde\sigma_n^{K^+ +K^-- 2K^0_s}(x,,y,z)=\nn
&&=\left\{\frac{1}{9} \,[4(d+\bar d) -(u+\bar u)]\left(1+\frac{\alpha_s}{2\pi}\otimes {\cal C}_{qq}\otimes\right) +\right.\nn
&&\qquad\left.+\frac{1}{3}\frac{\alpha_s}{2\pi} g\otimes {\cal C}_{gq}\otimes\right\}(D_u - D_d)^{K^+ +K^-},\label{c}
\ee

 Thus, due to SU(2)-invariance,  in all orders of QCD all three processes
 always measure the same NS combination of
fragmentation functions $(D_u-D_d)^{K^+ +K^-}$, whose evolution
does not involve the very poorly known gluon fragmentation
functions.

 The  difference of cross sections $K^++K^--2K^0_s$,
involving neutral kaons,  is essential in order to eliminate, due to SU(2) invariance, the $s+\bar s$-quark
parton densities and the gluon FF.

Note that the combinations of quark densities  in the above {\it do} have a singlet component and thus depend on $g(x)$, but
  that is not a problem.

 \subsection{LO approximation, $K^+ +K^-- 2K^0_s$}

 The LO expressions are particularly simple and obtained from (\ref{a})-(\ref{c}) with $\alpha_s=0$. They imply that SIDIS determines
$(D_u-D_d)^{K^++K^-}$ given $(u+\bar u)$ and $(d+\bar d)$ are known.

The difference $\tilde\sigma_p - \tilde\sigma_n$ is:
\be
&&\qquad (\tilde\sigma_p-\tilde\sigma_n)^{K^+ +K^-- 2K^0_s}(x,y,z)=\nn
&&=\frac{5}{9} \,[(u+\bar u) -(d+\bar d)]\, (D_u - D_d)^{K^+ +K^-},\label{p-nLO}
\ee
which is a non-singlet in both the PDs and the FFs. This implies that in its $Q^2$-evolution
and in all orders in QCD
it will always contain the same
NS combinations, convoluted with the corresponding Wilson coefficients when higher orders
are considered.

The fact that $e^+e^-$ and SIDIS measure the same combination $(D_u-D_d)^{K^++K^-}$ allows to combine $e^+e^-$ data  at $Q^2\simeq
m_Z^2$, where $Z^0$-exchange is the dominant contribution, with
SIDIS experiments  at $Q^2<< m_Z^2$ where
 $\gamma$-exchange dominates.
 For example one could test the relation
 \be
&&\frac{9\,d\tilde{\sigma}_p^{K^++K^--2K^0_s}(x,z,Q^2)}
{d\sigma_T^{K^++K^--2K^0_s}(z,m_Z^2)_{ \downarrow Q^2}}=\nn
&&\qquad\qquad\qquad =
\frac{[4(u+\bar u) -(d+\bar d)](x,Q^2)}{3\,\sigma_0\,(\hat e_u^2 -
\hat e_d^2)_{m_Z^2}}.\label{pe+e-}
 \ee
 Here $d\sigma_T^{K^++K^--2K^0_s}(z,m_Z^2)_{ \downarrow Q^2}$ denotes
that the data is measured at $m_Z^2$ and then evolved to $Q^2$
according to the DGLAP equations. This would be a test of LO, but
also a test of the factorization of SIDIS into parton densities times FFs.

 Tests for whether LO is a reasonable approximation for the SIDIS reactions
 can be made as follows. In LO one should have:

  1) for proton targets
\be
 &&\frac{\tilde\sigma_p^{K^+ +K^--
2K^0_s}(x,z)}{4(u+\bar u) -(d+\bar d)} = \mathrm{function \, of}
\, z \, \mathrm{only}\equiv  \nn
&&\qquad\equiv f_p(z) =
(D_u-D_d)^{K^++K^-}(z)\label{1}
\ee

2) for neutron targets
\be
  &&\frac{\tilde\sigma_n^{K^+
+K^-- 2K^0_s}(x,z)}{4(d+\bar d) -(u+\bar u)} = \mathrm{function \,
of} \, z \, \mathrm{ only}\equiv \nn
&& \qquad\equiv
f_n(z)=(D_u-D_d)^{K^++K^-}(z),\label{2}
\ee
where the PD's are
determined in LO, see for example~  \cite{Martin}.

3) and if measurements  for both proton and neutron targets are
available, then also
 \be
 \,\,f_p(z)=f_n(z)\label{3}
 \ee
  should
hold, as expected from (\ref{1}) and (\ref{2}).

The above LO-tests do not require knowledge of the FFs. Concerning
the measurement of FFs,  an attempt was made in   \cite{Kretzer_we}
to combine data on $e^+e^-$ and SIDIS. The evolution involved
there required an estimate of the gluon FF which induced quite
large errors. In the present case,
   we study the
 NS combination $(D_u - D_d)^{K^+ +K^-}$, which can be measured both  in $e^+e^-$ and SIDIS ,
  (\ref{a})-(\ref{c}),
   and whose evolution in $Q^2$ is straightforward since it does not involve the gluon FFs.

\subsection{NLO approximation, $K^+ +K^-- 2K^0_s$}

In higher orders of QCD the cross sections on $p$ and $n$ with
$K^+ +K^-- 2K^0_s$ depend on the gluon PD -- eqs. (\ref{b}) and (\ref{c}). The difference of
 the cross sections on proton and neutron  eliminates $g(x)$:
\be
&&\qquad (\tilde\sigma_p-\tilde\sigma_n)^{K^+ +K^-- 2K^0_s}(x,y,z)=\nn
&&=\frac{5}{9} \,[(u+\bar u) -(d+\bar d)]\left(1+\frac{\alpha_s}{2\pi}\otimes {\cal C}_{qq}\otimes\right)\nn
&&\qquad\qquad \times (D_u - D_d)^{K^+ +K^-}.\label{p-nNLO}
\ee
 and (\ref{p-nNLO}) determines $(D_u - D_d)^{K^+ +K^-}$ without the influence of even the gluon quarks or any other FF.
Note that $(u+\bar u) -(d+\bar d)$ is a NS  and thus $g(x)$ will not creep back through the $Q^2$-evolution.

Further, being a NS it would not be a problem to compare the two independent measurements: in $e^+e^-$ annihilation
at $Q^2\simeq m_Z^2$, eq. (\ref{a}), and in SIDIS at  $Q^2<< m_Z^2$, eq. (\ref{p-nNLO}).
They should give the same result, when evolved to the same $Q^2$ according to the DGLAP equations,
and thus present a test of the hypothesis that SIDIS is a product of the quark-production and quark-fragmentation processes.
This test would be independent of the gluon and strange PDs or any other FFs and hold in any order in QCD.

Having thus determined $(D_u - D_d)^{K^+ +K^-}$ one may proceed to determine the gluon PD, without the uncertainties of
$s+\bar s$, measuring the sum of the same cross sections on $p$ and $n$:
\be
&&\qquad (\tilde\sigma_p+\tilde\sigma_n)^{K^+ +K^-- 2K^0_s}(x,,y,z)=\nn
&&=\frac{1}{3} \left\{[(u+\bar u) +(d+\bar d)]\left(1+\frac{\alpha_s}{2\pi}\otimes {\cal C}_{qq}\otimes\right) +\right.\nn
&&\qquad\left.+2\frac{\alpha_s}{2\pi} g\otimes {\cal C}_{gq}\otimes\right\}(D_u - D_d)^{K^+ +K^-},\label{p+n}
\ee

%\vspace{0.5cm}

%%%%%%%%%%%%%%%%%%%%%%%%%%%%%%%%%%%%%%%%%%%%%%%%%%%%%%%%%%%%%%%%%%%%%%%%%%%%%%%%%%%%%%%%%%%%%%%%%%
\section{The combination $K^+ +K^-+ 2K^0_s $}
\label{sec:9}
%%%%%%%%%%%%%%%%%%%%%%%%%%%%%%%%%%%%%%%%%%%%%%%%%%%%%%%%%%%%%%%%%%%%%%%%%%%%%%%%%%%%%%%%%%%%%%%%%%%%%%%

The general expressions in NLO are rather lengthy, so we begin by discussing the LO case which already exhibits the main properties.
For brevity we use the notation $(K)\equiv K^+ +K^-+ 2K^0_s$.
\subsection{LO approximation, $K^+ +K^-+ 2K^0_s$}

In LO we have

  for $e^+e^-$:
\be
d\sigma^{(K)}_T(z)&=& 3\,\sigma_0\left[(\hat e^2_u + \hat e^2_d
)_{m_Z^2} (D_u + D_d)^{K^+ +K^-}+\right. \nn
&&\qquad \left.+ 2 \,\hat e^2_d D_s^{K^+
+K^-}\right],\label{e+e-K}\\
 d\sigma^{(K)}_T &\equiv&
d\sigma^{K^+}_T\,+d\sigma^{K^-}_T\,+2\,d\sigma^{K^0_s}_T
\nee

and for SIDIS:
\be
    &&\qquad \tilde\sigma_p^{(K)} (x,z,Q^2) =\frac{1}{9}\left[(4(u+\bar u)+\right. \nn
   && \left.+ (d+\bar d))
 (D_u+D_d)^{K^++K^-} + 2(s+\bar s)D_s^{K^++K^-}\right]\label{p}\\
  &&\qquad \tilde\sigma_n^{(K)} (x,z,Q^2)
=\frac{1}{9}\left[(4(d+\bar d)+\right.\nn
&&\left.+ (u+\bar u)) (D_u+D_d)^{K^++K^-}+
2(s+\bar s)D_s^{K^++K^-}\right].\label{n}
 \ee

Eqs. (\ref{e+e-K}) - (\ref{n}) imply that due to SU(2) invariance, the three cross sections
$d\sigma^{(K)}_T$, $\tilde\sigma_p^{(K)}$ and $\tilde\sigma_n^{(K)}$ always measure  only two
combinations of FFs:
$(D_u+D_d)^{K^++K^-}$ and $D_s^{K^++K^-}$. Note that, as this is  a property of SU(2)-symmetry,
it will hold in all orders of QCD,
only the gluon FF will enter in addition in higher orders.

 From (\ref{e+e-K}) - (\ref{n}) it follows that in LO we have three measurements for  two unknown quantities
$(D_u+D_d)^{K^++K^-}$ and $D_s^{K^++K^-}$. This implies
in particular that measurements of $K^++K^--2 K_s^0$ and $K^++K^-+2 K_s^0$ in SIDIS -- eqs. (\ref{b}), (\ref{c}), (\ref{p}) and (\ref{n}),
are already enough to determine $(D_u\pm D_d)^{K^++K^-}$ and $D_s^{K^++K^-}$ and
it is not necessary to use data from $e^+e^-$  performed at very different $Q^2$.

The difference
$\tilde\sigma_p^{(K)}-\tilde\sigma_n^{(K)}$ determines $(D_u+D_d)^{K^++K^-}$ only through the NS
combination $(u+\bar u)-(d+\bar d)$:
\be
\tilde\sigma_p^{(K)}-\tilde\sigma_n^{(K)}=\frac{1}{3}[(u+\bar u)-(d+\bar d)]\,(D_u+D_d)^{K^++K^-}.
\ee
Once we have thus determined $(D_u+D_d)^{K^++K^-}$, we can use  $\tilde\sigma_{p,n}^{(K)}$
(or equivalently their sum $\tilde\sigma_p^{(K)}+\tilde\sigma_n^{(K)}$)
 to obtain $D_s^{K^++K^-}$.

 Only in LO are SIDIS measurements are enough to determine
$D_{u,d,s}^{K^++K^-}$.  It is thus important to have reliable tests of LO approximation.
It's an advantage that using the same expressions (\ref{p}) - (\ref{n})
one can form  possible tests of the LO in these processes.

 1) In LO  we have:
\be
&& \frac{3\,(\tilde\sigma_p-\tilde\sigma_n)^{K^+ +K^-+
2K^0_s}(x,z)}{\left(u+\bar u -(d+\bar d)\right)(x)}=
\mathrm{function \, of} \, z \, \mathrm{ only} =\nn
&&\qquad =(D_u+D_d)^{K^++K^-}(z). \label{5}
\ee

2)  If $K_s^0$ are not measured, LO would be a good approximation
if:
\be
 && \frac{9(\tilde\sigma_p-\tilde\sigma_n)^{K^+
+K^-}(x,z)}{\left(u+\bar u -(d+\bar d)\right)(x)}=
\mathrm{function \, of} \, z \, \mathrm{ only} =\nn
&&\qquad =
(4D_u-D_d)^{K^++K^-}(z),\label{6}
\ee
 i.e. only the combination of FFs on the r.h.s. is different from (\ref{5}).

 For neither of these tests is a knowledge of the FFs  necessary and they should proceed the extraction of
$D_q^{K^++K^-}$.

\subsection{NLO approximation, $K^+ +K^-+ 2K^0_s$}

As mentioned, in NLO the three cross sections $d\sigma^{(K)}_T$, $\tilde\sigma_p^{(K)}$ and $\tilde\sigma_n^{(K)}$
measure different combinations of the three unknown FFs:
\be
(D_u+D_d)^{K^++K^-},\,\, D_s^{K^++K^-}\quad {\rm and}\quad D_g^{K^++K^-}.\label{FFs}
\ee
(The general expressions for NLO are rather lengthy, so we present bellow only those relevant for
our discussion.) This implies that in NLO, contrary to LO, both $e^+e^-$ and SIDIS measurements are needed to determine
(\ref{FFs}). Combined with measurements of $K^++K^--2K_s^0$, eqs. (\ref{a}) - (\ref{c}), we have enough measurements
to determine all kaon FFs: $(D_u\pm D_d)^{K^++K^-}$, $D_{s}^{K^++K^-}$ and $D_{g}^{K^++K^-}$.

Solely from SIDIS, and without the influence of the strange and gluon PDs, in NLO one can determine
$D_{u\pm d}^{K^++K^-}$ and $D_g^{K^++K^-}$. The difference of  $(\tilde\sigma_p-\tilde\sigma_n)^{(K)}$  determines
 a combination of $(D_u + D_d)^{K^++K^-}$ and $D_g^{K^++K^-}$, where the PDs enter only as a common factor in the combination
 $(u+\bar u) -(d+\bar d)$:
 \be
&&\qquad (\tilde\sigma_p-\tilde\sigma_n)^{K^+ +K^-+ 2K^0_s}(x,,y,z)=\nn
&&=\frac{1}{3} [(u+\bar u) -(d+\bar d)]\left\{[1+\frac{\alpha_s}{2\pi}\otimes {\cal C}_{qq}\otimes ]\times \right.\nn
&&\left.\times (D_u + D_d)^{K^+ +K^-} +2\frac{\alpha_s}{2\pi} \otimes {\cal C}_{qg}\otimes D_g^{K^+ +K^-}\right\}.\label{+Ks}
 \ee
As these FFs are not NS and thus have a different $Q^2$-evolution, the above equation would provide
information on both $(D_u + D_d)^{K^++K^-}$ and $D_g^{K^++K^-}$. Further, combined with measurements on
$(D_u - D_d)^{K^++K^-}$ from (\ref{p-nNLO}) one can determine $(D_u\pm D_d)^{K^++K^-}$ and $D_g^{K^++K^-}$ in NLO solely in SIDIS
and they will depend on the parton densities only through the combination  $(u+\bar u) -(d+\bar d)$.

 Further one can combine the
measurements of $D_{u\pm d}^{K^++K^-}$ and $D_g^{K^++K^-}$ with measurements
of $e^+e^-$ annihilation or $p + n$ SIDIS cross section to determine $D_s^{K^++K^-}$.
 Especially useful would be $e^+e^-$ annihilation
 where $D_s^{K^++K^-}$ is not multiplied by the small quantity $(s+\bar s)$:
\be
&&d\sigma^{(K)}_T(z)= 3\,\sigma_0\left\{\left((\hat e^2_u + \hat e^2_d
)_{m_Z^2} (D_u + D_d)^{K^+ +K^-}+\right.\right. \nn
&& \qquad\,\left.+ 2 \,\hat e^2_d D_s^{K^+
+K^-}\right)\,\left[1+\frac{\alpha_s}{2\pi}\otimes C_F\,(c_T^q+c_L^q)\right]+\nn
&&\qquad\,\left.+2\frac{\alpha_s}{2\pi}(\hat e^2_u + 2\hat e^2_d)_{m_Z^2}\otimes C_F(c_T^g+c_L^g) D_g^{K^+ +K^-}\right\}
\nee
The advantage is  that in this way neither the strange nor the gluon parton densities influence the determination of the kaon FFs.
\vspace{.5cm}

In summary, if in addition to  charged $K^\pm$ also  neutral
$K^0_s$ are measured, we showed that in LO  all FFs
$D_{u,d,s}^{K^++K^-}$ can be determined  solely from SIDIS, i.e. it is not necessary
to use data from $e^+e^-$- performed at very different $Q^2$.
  In NLO $e^+e^-$ data should be included, as well,
  and then all FFs can be determined without the influence of the strange and gluon PDs.
  The
non-singlet $(D_u-D_d)^{K^++K^-}$ can be singled out in both $e^+e^-$ and SIDIS.
Since  comparing the two measurements at different $Q^2$ is straightforward,
one can test the factorization of the SIDIS cross section into parton densities and fragmentation functions
both in LO and NLO.

%%%%%%%%%%%%%%%%%%%%%%%%%%%%%%%%%%%%%%%%%%%%%%%%%%%%%%%%%%%%%%%%%%%%%%%%%%%%%%%%%%%%%%%%%%%%%%%%%%
\section{Conclusions}
\label{sec:10}
%%%%%%%%%%%%%%%%%%%%%%%%%%%%%%%%%%%%%%%%%%%%%%%%%%%%%%%%%%%%%%%%%%%%%%%%%%%%%%%%%%%%%%%%%%%%%%%%%%%%%%%

The paper considers the possibilities to obtain the kaon FFs in $e^+e^-$ annihilation and SIDIS.
 It consists of two parts. In the first part
we have considered possible tests for $s-\bar s=0$ and
$D_d^{K^+-K^-}= 0$ in unpolarized SIDIS with final charged
$K^\pm$, both in LO and NLO of QCD.

In the second part we show that, if in addition
to $K^\pm$ also the neutral  $K_s^0$ are measured 1) in LO the kaon FFs can be obtained solely from SIDIS, and
2) in NLO  the combined data of the total cross section in
$e^+e^-$ annihilation in addition to SIDIS is also needed; then the  FFs
can be determined without the uncertainties of the strange and gluon PDs.
Different possibilities to test the LO approximation in unpolarized SIDIS are discussed and in all
proposed tests no knowledge of the fragmentation functions is  necessary.
We show that, in all orders of QCD, the non-singlet combination $(D_u-D_d)^{K^++K^-}$
can be measured directly both in $e^+e^-$ and in SIDIS without any influence of the strange and gluon parton densities
or any other FFs.
Comparing the measurements in $e^+e^-$ and SIDIS allows tests of the factorization of SIDIS
into parton densities and fragmentation functions in any order in QCD.

In our approach we consider the sum and difference of  cross
sections for  hadron $h$ and its C-conjugate $\bar h$. The  cross
section differences , $h-\bar h$, are NS and, both   their
$Q^2$-evolution and NLO corrections in QCD are straightforward,
since they  don't mix with other PDs or FFs. But they involve
poorly known quantities such as the non-singlets $s-\bar s$ and
$D_d^{K^+-K^-}$, and  we suggest some tests
 for these quantities. Quite the opposite is true when the sum of cross sections
$h+\bar h$ is considered. In this case the $Q^2$-evolution and NLO
corrections involve the poorly known gluon FFs, but the cross
sections contain the best known combinations of PDs  $q+\bar q$,
measured in DIS, and $D_q^{h^++h^-}$ measured in $e^+e$.

We have tried to exploit some of the advantages of both types of
combinations of data. Note that though we often consider difference asymmetries,
the quantities that they determine are not small and thus, we hope,  measurable.

We want to add few remarks on the measurability of the discussed asymmetries. In general,
 these are difference asymmetries and high precision measurements are required.
 In addition, the data should be presented in bins in both $x$ and $z$.
% We can distinguish schematically two types of asymmetries: with charged
% kaons only and with both charged and neutral kaons.
Quite recently such binning was done in \cite{PHD}
for the very precise data of the HERMES collaboration in DESY on $K^\pm$-production in semi-inclusive DIS
  on Proton and Deuterium targets .
  These  results  show that for $0,350\leq z\leq 0,450$ and for $0,450 \leq z\leq 0,600$
  in the $x$-interval $0,023 \leq x\leq 0,300$
 the accuracy of the data allows to form the differences $(\sigma_p+\sigma_n)^{K^+-K^-}$ and
 $(\sigma_p-\sigma_n)^{K^+-K^-}$ with errors not bigger than ~  7-13\% and ~10-15\% respectively. Having these
 cross sections, given that $u_V$ and $d_V$ are well known, one can form the ratios
 $R_+$ and $R_-$ with these precisions. Then, if we do not obtain an acceptable fit to
$R_+(x,z_0)$ which is independent of x, then $s-\bar s=0$ is not a good
approximation. This conclusion assumes   the success of the LO-test
involving $R_-(x,z_0)$, and is independent of our knowledge of the
FFs.

If however, an acceptable x-independent fit to $R_+(x,z_0)$ is obtained, then
the precision of this fit will put limits on $(s-\bar s)D_s^{K^+-K^-}$. Using
these limits in the expression for $R_+ -R_-$, and comparing it
with experiment at the same values $z_0$
will then put limits on $D_d^{K^+-K^-}$.

If we work in NLO and we do not succeed to obtain an acceptable fit for (\ref{NLO1}) and (\ref{NLO2}) with the same $D(z)$, then
 $s-\bar s\simeq 0$ and $D_d^{K^+-K^-}\simeq 0$ cannot hold simultaneously, at least one of these assumptions fails.

\section*{Acknowledgements}
This work was supported by a Royal Society International Joint
Project Grant.


\begin{thebibliography}{99}
\bibitem{Kretzer_we}
S. Kretzer, E. Leader, E. Christova,   Eur.Phys.J. {\textbf  C22}
(2001) 269-276
\bibitem{deFlorian} D. de Florian, G.A.Navaro and R. Sassot,  Phys. Rev. {\textbf D71} (2005) 094018
\bibitem{Dubna05}
E. Christova, E. Leader, {\textit  Proceedings of
 XI-th workshop on high energy physics, Dubna-SPIN-2005} (Russia), pp. 17-21 (Preprint: hep-ph/0512075)
 \bibitem{Vogelsang}D.de Florian, M. Stratmann and W. Vogelsang, Phys. Rev. {\textbf D57} (1998) 5811
 \bibitem{Kretzer} S. Kretzer, Phys. Rev. {\textbf D62} (2000) 054001
 \bibitem{Martin} A.D Martin, R.G. Roberts,W.J.Stirling and R.S.Thorne,  Phys. Lett.
 {\textbf  B531} (2002) 216
 \bibitem{PHD} Achim Hillenbrand, Measurement and Simulation of the Fragmentation
Process at HERMES, Ph.D. theses, DESY-2005
 \end{thebibliography}
\end{document}